\documentclass[manuscript]{aastex}
\usepackage{ulem}
\usepackage{lineno}

\slugcomment{Not to appear in Nonlearned J., 45.}

\shorttitle{Reviewing Emission and Profiles of RX J1713.7-3946}
\shortauthors{Tang et al.}

\begin{document}


\title{Radial Profiles of Non-thermal Emission from Supernova Remnant RX J1713.7-3946}


\author{Yunyong Tang}
\affil{School of Physical Science and Technology, Kunming University, Kunming 650214, China
\\Key Laboratory of Dark Matter and Space Astronomy, Purple Mountain Observatory, Chinese Academy of Sciences, Nanjing 210034, China tangyunyong888@163.com}
\author{Siming Liu}
\affil{Key Laboratory of Dark Matter and Space Astronomy, Purple Mountain Observatory, Chinese Academy of Sciences, Nanjing 210034, China liusm@pmo.ac.cn}



\begin{abstract}
Supernova remnant RX J1713.7-3946 (also named as G347.3-0.5) has exhibited largest surface brightness, detailed spectral and shell-type morphology, it is one of the brightest TeV sources. The recent H.E.S.S. observation of RX J1713.7-3946 revealed \textbf{a} broken power-law spectrum of GeV-TeV gamma-ray spectrum and more extended gamma-ray spatial radial profile than that in the X-ray band. Based on the diffusion shock acceleration model, we solve spherically symmetric hydrodynamic equations and transport equations of particles, and investigate multi-band non-thermal emission of RX J1713.7-3946 and radial profiles of its surface brightness for two selected zones in the leptonic scenario for the $\gamma$-ray emission. We found (1) the diffusion coefficient has a weak energy-dependent, and the Kolmogorov type is favored; (2) the magnetic field strength could vary linearly or nonlinearly with radius for different surrounding environments because of possible turbulence in shock downstream region, and a compressional amplification is likely to exist at the shock front; (3) the non-thermal photons from radio to X-ray bands are dominated by synchrotron emission from relativistic electrons, if the GeV-TeV gamma-rays are produced by inverse Compton scattering from these electrons interacting with the background photons, then the X-ray and gamma-ray radial profiles can be reproduced except for the more extended $\gamma$-ray emission.
\end{abstract}


\keywords{Theoretical models, Radiative processes, Non-thermal radiation sources, Gamma-rays, Supernova remnants}



\section{Introduction}           
Supernova remnants (SNRs) are considered as well-known accelerators of Galactic cosmic rays, while cosmic rays below the knee energy ( $\sim10^{15}$ eV ) are thought likely to come from these SNRs \citep{Blasi2013}. Coincidentally, cosmic-ray particles could be effectively accelerated to such high energy by the diffusive shock acceleration (DSA), which results in a power-law distribution of particles \citep[e.g.][]{Bell1978,Blandford1987,Malkov2001}. In view of the excellent observation results of spectral and morphological features of young SNRs, the synchrotron X-ray emission observed from young SNRs manifests a presence of multi-TeV energy electrons, which naturally confirms the deductions of the DSA. Based on the DSA mechanism, the multi-band emission from radio to gamma-ray band can be from electron synchrotron radiations, inverse Compton scatterings (ICSs) and non-thermal bremsstrahlung emissions (leptonic scenario) or gamma-ray emission from  $\pi^0$-decay process with collisions between cosmic rays and background protons (hadronic scenario). However, it is still controversial that the gamma-rays are produced by either a hadronic or leptonic (or hybrid) procedure. Up to now, many details of the DSA mechanism are also still unclear, such as the acceleration and injection efficiencies of accelerated particles, the maximum energy of cosmic rays accelerated by supernova shocks, as well as the roles of particle diffusion and magnetic field strength.

The SNR RX J1713.7-3946 is a well-known source for the DSA mechanism study on account of detailed observations of radio, X-ray, and gamma-ray emission components \citep[e.g.][]{Acero2009,Koyama1997,Cassam2004,Tanaka2008,Abdo2011,Aharonian2004,Aharonian2006,Aharonian2007,H.E.S.S.Collaboration2018}, it should be an efficient cosmic ray accelerator. \citet{Pfeffermann1996} first discovered the RX J1713.7-3946 by ROSAT X-ray observations, and then \citet{Wang1997} judged that the RX J1713.7-3946 was the remnant of the AD393 guest star according to the ROSAT observation, the historical records as well as the visual position, distance, and age. It can be inferred that its age should be $\sim$ 1625 yr. A distance of 1 kpc was derived by \citet{Koyama1997} by means of the X-ray measurements of the column density toward this source, and it also agrees with the estimated result from new high-resolution CO mm-wave observations \citep{Fukui2003}.

ASCA observations indicated that the X-ray emission from RX J1713.7-3946 is predominantly non-thermal emission, the line emission from the SNR interior was not detected, and an upper limit on the mean density around the remnant of less than 0.3 cm$^{-3}$  was set with considering the lack of thermal emission \citep{Slane1999}. Subsequently, a stringent constraint on the post-shock gas density is also given with its value of $\leq$ 0.8 cm$^{-3}$ in view of the absence of thermal X-ray emission \citep{Tanaka2008}. ROSAT and ASCA images is resolved with Chandra into bright filaments and fainter diffuse emission, which shows good correspondence with the radio morphological structure, and provides strong proof that the same population of electrons is responsible for the synchrotron emission in radio and X-ray bands in northwestern region of the remnant \citep{Lazendic2004}. In addition, the observations of Chandra, XMM-Newton, and Suzaku \citep{Uchiyama2003,Cassam2004,Hiraga2005,Takahashi2008,Tanaka2008} also proclaimed that the X-ray emission is predominantly from the synchrotron component without an evidence of a thermal X-ray component. Suzaku measurements have stated clearly a power law spectrum with a photon index of $\sim$ 2 and a smooth cutoff around a few keV, the hard X-ray spectrum from Suzaku HXD up to 40 keV has also been well described by a power-law form with the photon index of $\sim$ 3, which is steeper than that measured with energies below $\sim$ 10  keV. Recently, the INTEGRAL 17-120 keV spectrum of RX J1713.7-3946 is given with a power-law continuum with the photon index of $\sim$ 3, that is also softer than the value below $\sim$ 10 keV, it is obvious that the spectrum becomes gradually steeper with increasing energy \citep{Kuznetsova2019}. \citet{Tsuji2019} exhibited the spatially resolved non-thermal X-ray emission up to 20 keV with \emph{NuSTAR}, the photon index is 2.15 and the cutoff energy is 18.8 keV in their model, they founded that the cutoff energy seem to be variable from 0.6 to 1.9 keV, to some extent, the cutoff shape can also give us some indication about the diffusion type. In addition, \cite{Katsuda2015} reported the first evidence for thermal X-ray line emission from RX J1713.7-3946, these lines can be explained as the thermal emission from reverse-shocked supernova ejecta rather than swept-up interstellar medium (ISM) or circumstellar medium (CSM), the progenitor of this remnant was a relatively low-mass star($\le$ 20 $M_\odot$), and RX J1713.7-3946 is inferred as a result of an SN Ib/c whose progenitor was a member of an interacting binary.

The observed gamma-ray energy spectrum of RX J1713.7-3946 have covered five orders of magnitude in energy with combining Fermi-LAT and H.E.S.S. observations. RX J1713.7-3946 was firstly detected in very-high-energy (VHE) gamma rays by the CANGAROO collaboration \citep[e.g.][]{Muraishi2000,Enomoto2002}. Afterwards, H.E.S.S. Collaboration confirmed that the VHE gamma-ray radiation comes from the SNR shell and afforded the resolved VHE gamma-ray image, the later observations further provided a lot of detailed information about morphology, radial profiles and VHE gamma-ray energy spectrum \citep[e.g.][]{Aharonian2004,Aharonian2006,Aharonian2007,H.E.S.S.Collaboration2018}. In GeV energy range, \citet{Abdo2011} investigated the gamma-ray emission from RX J1713.7-3946 with the Fermi-LAT, it is shown that the spectral index in this band is very hard with a photon index of 1.5 $\pm$ 0.1 which is well in agreement with leptonic emission scenarios, in other words, the gamma-ray emission could be dominated by the ICS of ambient lower energy photons by the relativistic electrons. The good correlation between the gamma-ray and X-ray image is clearly presented by means of the XMM-Newton hard X-ray contours overlaid on the H.E.S.S. gamma-ray excess image \citep{H.E.S.S.Collaboration2018}. Through the deep H.E.S.S. exposure and detailed spectral and morphological studies, a significant result is shown that the TeV gamma-ray emission extends beyond the X-ray emission associated with the SNR shell. Therefore the correlation studies of the X-ray and gamma-ray emission will play an important role in the radiation origin of RX J1713.7-3946.

Up to now, several theoretical models have been constructed to explain the multi-band emission from RX J1713.7-3946, such as leptonic scenarios \citep[e.g.][]{Liu2008,Fan2010a,Abdo2011,Li2011,Yuan2011,Finke2012,Ellison2012,Lee2012,Yang2013,Zhang2019}, hadronic scenarios \citep[e.g.][]{Berezhko2008,Berezhko2010,Fang2009,Fang2011,Inoue2012,Gabici2014,Federici2015}, and hybrid scenarios (involving lepton and hadron) \citep[e.g.][]{Zirakashvili2010,Zeng2019}, and some scenarios have taken nonlinear effects into account. For the sake of simplicity, the pressure of energetic particles is neglected in our calculation, we only consider the test particle approximation, moreover, here we are in favor of the leptonic scenario on account of the GeV harder sepctrum and the good correlation between the X-ray and gamma-ray image of RX J1713.7-3946. Nevertheless, the emission origin of RX J1713.7-3946 is still debatable. As mentioned above, the diffusion process and the magnetic field strength are also uncertain in the DSA mechanism, the diffusion process is regarded as energy-dependent in some cases, such as Bohm type, Kraichnan type and Kolmogorov type \citep[e.g.][]{Bhattacharjee2000,Burlaga2015}, however, the diffusion could be also dominated by turbulent mixture \citep[e.g.][]{Bykov1993,Fan2010b,Zhang2017}. The magnetic field strength is very complex especially in the shock downstream region, which can be amplifed by several ways, such as compressional amplification \citep{Iapichino2012}, amplification due to current-driven instabilities \citep{Bell2004} and turbulent amplification \citep[e.g.][]{Giacalone2007,Guo2012,Ji2016,Xu2016,Xu2017}. In our model, the diffusion coefficient and the magnetic field strength will be determined by fitting the observed data. The detailed observations provide us with a great opportunity to study RX J1713.7-3946, especially, good correlation has been found between the X-ray and gamma-ray image of RX J1713.7-3946, it seems to make sense for further research on the emission scenario and radial profiles of RX J1713.7-3946.

In this paper, we apply the DSA model with test particle approximation to research the multi-band no-thermal emission spectra and radial profiles of RX J1713.7-3946, and only investigate the leptonic scenario for studying the correlation between the X-ray and gamma-ray emissions. The structure of this paper is organized as follows, in Section 2, we describe briefly the theoretical model and carry out relevant parameter analysis, the model is applied to the RX J1713.7-3946 and the corresponding results are shown in Section 3, finally we give our conclusion and discussion in Section 4.

\section{Model and parameter analysis}
\label{sect:Mod}
\subsection{The model}
After supernova explosion, the forward shock and reverse shock are generated by supersonicly moving supernova ejecta. The forward shock propagates in the circumstellar medium, as well as the reverse shock propagates in the ejected gas. A numerical kinetic approach about nonlinear DSA have been developed by \citet{Zirakashvili2010}, and they put forward numerical solution of spherically symmetric hydrodynamic equations by combining with the energetic particle transport and acceleration by means of the forward shock and reverse shock. It is generally assumed that some part of thermal particles is injected at the shock fronts into acceleration.

In the model, the gas density $\rho(r, t)$, gas velocity $u(r, t)$, gas pressure $P_g(r, t)$ are described in the hydrodynamical equations, as shown below
\begin{equation}
\frac {\partial \rho }{\partial t}=-\frac {1}{r^2}\frac {\partial }{\partial r}r^2u\rho
\end{equation}
\begin{equation}
\frac {\partial u}{\partial t}=-u\frac {\partial u}{\partial r}-\frac {1}{\rho }
\left( \frac {\partial P_g}{\partial r}+\frac {\partial P_c}{\partial r}\right)
\end{equation}
\begin{equation}
\frac {\partial P_g}{\partial t}=-u\frac {\partial P_g}{\partial r}
-\frac {\gamma _gP_g}{r^2}\frac {\partial r^2u}{\partial r}
-(\gamma _g-1)(w-u)\frac {\partial P_c}{\partial r},
\end{equation}
and the cosmic-ray proton momentum distribution $N(r, t, p)$ in the spherically symmetrical case are given by\[
\frac {\partial N}{\partial t}=\frac {1}{r^2}\frac {\partial }{\partial r}r^2D(r,t,p)
\frac {\partial N}{\partial r}
-w\frac {\partial N}{\partial r}
\]
\[+\frac {\partial N}{\partial p}
\frac {p}{3r^2}\frac {\partial r^2w}{\partial r}-\frac{1}{p^2}\frac {\partial}{\partial p}p^2b(p)N
\]
\[
+\frac {\eta ^f\delta (p-p_{f})}{4\pi p^2_{f}m}\rho (R_f+0,t)(\dot{R}_f-u(R_f+0,t))\delta (r-R_f(t))
\]
\begin{equation}
+\frac {\eta ^b\delta (p-p_{b})}{4\pi p^2_{b}m}\rho (R_b-0,t)(u(R_b-0,t)-\dot{R}_b)\delta (r-R_b(t))
\end{equation}
where $P_c = 4\pi\int{p^2dpvpN/3}$ is the cosmic ray pressure, $w(r, t)$ is the advective velocity of cosmic rays,
$\gamma_g=5/3$ is the adiabatic index of the gas, $D(r, t, p)$ is the cosmic ray diffusion coefficient, and $b(p)=-dp/dt$ is energy-loss rate of particles. If the diffusive streaming of cosmic ray induced the generation of magneto-hydrodynamical waves, in which the energetic particles are scattered to result in the difference between the cosmic ray advective velocity $w$ and the gas velocity $u$. Their difference in the value of the radial component of Alf\'en velocity $(V_A = B/\sqrt{4\pi\rho})$ is $w = u + \xi_{A}V_A/p^3$ in the isotropic random magnetic field $B$ \citep{Zirakashvili2012}.
The parameter $\xi_A$ represents the relative direction of fluid velocity to Alf\'en velocity with $\xi_A = -1$ in upstream of forward shock and $\xi_A = 1$ in upstream of reverse shock respectively, the value of $\xi_A$ is zero in the other region (e.g. the downstream for forward and reverse shocks ). Two last terms in Eq. (4) correspond to the injection of thermal protons with
momenta $p = p_f$ , $p = p_b$ and mass $m$ at the fronts of the forward and reverse shocks at
$r = R_f(t)$ and $r = R_b(t)$ respectively. The dimensionless parameters $\eta_f$ and $\eta_b$ determine
the injection efficiency. Hereafter, the indexes $f$ and $b$ for quantities are represented the
forward and reverse shock respectively.

In our study, we only compute the contribution from energetic electrons, and the electrons are regarded as test particles, in other words, the pressure of electrons is neglected (i.e. $P_c = 0$). In leptonic scenario, the electrons could lose energy through some processes such as synchrotron emission, ICSs, bremsstrahlung emission, ionization losses, and so on, the corresponding energy-loss time-scales of above-mentioned processes can be respectively expressed as $t_{\rm syn}=1.3\times10^5(E/{\rm TeV})^{-1}(B/10~\mu{\rm G})^{-2}~{\rm yr}$, $t_{\rm IC}\approx t_{\rm syn}U_{\rm B}/U_{\rm ph}$, $t_{\rm Brem}=3.3\times10^5(n_{\rm 0}/100~{\rm cm}^{-3})^{-1}~{\rm yr}$, $t_{\rm ion}=1.9\times10^4(n_{\rm 0}/100~{\rm cm}^{-3})^{-1}\gamma/(3{\rm ln}{\gamma}+18.8)~{\rm yr}$, here $E$ and $\gamma$ are the electron energy and Lorenz factor, the magnetic energy density $U_{\rm B}=B^2/8\pi$, $U_{\rm ph}$ is the energy density of background photon field and $n_{\rm 0}$ is the surrounding medium density\citep{Ginzburg1964}. The synchrotron loss time of electrons with their energies $\sim$TeV is shorter with our model parameters, the synchrotron energy-loss dominate over the other processes, so we consider only synchrotron energy loss process with the energy-loss rate $b(p)=4e^4pB^2/(9m_e^3c^6)$ in our calculations.
For simplify, the cosmic ray drift velocity is considered being consistent with the gas velocity (i.e. $w = u$). The equation for the quasi-isotropic cosmic ray momentum distribution $N(r, t, p)$ were numerically solved in the spherically symmetrical case by using a finite-difference method. The radiation energy spectrum and spatial profile of the supernova remnant are closely related to the particle injection and diffusion process during supernova-remnant evolution, but some problems about the particle injection and diffusion are still unresolved, it seems to be more difficult to solve these problems for the reverse shock, so the distributions from the reverse shock are ignored in our work, i.e., particle accelerations were only considered in forward shock. The numerical procedure of solving Equation (1)-(4) was described in detail in \citet{Zirakashvili2012}.

The magnetic field plays no dynamical role in our calculation, here the evolution process of magnetic field strength can not be modeled at shock, because no magneto-hydrodynamic process is involved. Commonly, the shocked ejecta and interstellar gas are separated by the contact discontinuity at $r = R_c$. For forward shock ($r > R_c$), post-shock magnetic fields could be amplified by means of some above-mentioned mechanisms, such as compression, current-driven instabilities and turbulent dynamo. Through these amplification mechanisms, the average downstream magnetic fieldis can increase significantly\citep[e.g.][]{Iapichino2012,Bell2004,Giacalone2007,Guo2012,Ji2016,Xu2016,Xu2017}, however, the detailed magnetic field structure is still controversial, here it is just phenomenologically assumed. In our work, because of smaller dynamical effects of the magnetic fields in upstream of the forward shock when $r > R_f$, the magnetic field is fixed to a value of $B_0$ that may be slightly larger than the interstellar magnetic field of $\sim 3~\mu$G. In downstream of the forward shock, we simply assume that the magnetic field strength satisfies the following relationship as
\begin{equation}
B(r,t)=\left\{ \begin{array}{lll}
B_0+\frac{B_m-B_f}{(R_p-R_c)^\alpha}{(r-R_c)^\alpha}, \ R_c<r<R_p,
\\
B_f +\frac{B_m-B_f}{(R_p-R_f)^\alpha}{(r-R_f)^\alpha}, \ R_p<r\le R_f,
\\
B_0, \ r > R_f
\end{array} \right.
\end{equation}
here $R_p=R_c+\xi(R_f-R_c)$ represents the position where the maximum magnetic field strength $B_m$ occurs in downstream region of the forward shock, $\xi$ and $\alpha$ are two parameters that determine the distribution of the magnetic field. $B_0$ and $B_f$ are magnetic field strength in upstream of the forward shock and at the shock front (at $r = R_f$) respectively. Although we do not considered the contribution of the reverse shock, the magnetic field strength in reverse shock could also exist as considered by \citet{Zirakashvili2012}, here we simply set $B=3~\mu$G within the reverse shock.

We take into account energy-dependent diffusion, on the basis of multi-band observation data, and further judge whether the diffusion coefficient is strongly related to energy or not. The diffusion coefficients which depend on energy of cosmic ray particles can be expressed as
\begin{equation}
D(E)=10^{28}\chi (\frac{E}{10~\rm{GeV}})^\delta
\label{dif2}
\end{equation}
where the accelerated particle energy is $E$, $\chi$ is the correction factor and $\delta$ is the index which depends on the turbulence spectrum of the magnetic field, just as $\delta=0.0$ for energy-independent diffusion, $\delta=1/3$ for Kolmogorov type, $\delta=1/2$ for Kraichnan type, $\delta=1$ for Bohm type \citep[e.g.][]{Berezinskii1990,Bhattacharjee2000,Fujita2009}.

Based on the DSA model, the hydrodynamic and particle propagation equations are solved by means of combining the distribution of magnetic field strength and energy-dependent diffusion. Subsequently, the spectra of accelerated particles and photon spectra during the whole evolution of RX J1713.7-3946 can be produced. After the projection effect is taken into account, we furter caculate the radial profiles of brightness distributions of X-ray and gamma-ray emissions from RX J1713.7-3946. The photon emissions from radio to X-ray band are dominated by the synchrotron emission from accelerated electrons, the GeV-TeV gamma-rays are produced by the ICS of high energy electrons. For calculation of radiation, one can refer to some related formulas \citep[e.g.][]{Blumenthal1970,Rybicki1979,Zhang2007}.

\subsection{The parameter analysis}
\label{sect:Para}
In this section, we apply the model to the RX J1713.7-3946, and some appropriate parameters will be picked out to fit the multi-band photon emission spectrum and radial radiation profile. In consideration of existing theories and observations as described in Section 1, the age of the RX J1713.7-3946 is fixed at $t_{\rm age}=1625 ~\rm yr$, and its distance $d$ is equal to 1 kpc. After the supernova explosion, the supernova ejecta \textbf{is} thought to have some velocity distribution as \citep{Zirakashvili2012}
\begin{equation}
P(V)=\frac{3(k-3)}{4\pi k}\left\{ \begin{array}{ll}
1, \ V<V_{\rm ej},
\\
(V/V_{\rm ej})^{-k}, \ V>V_{\rm ej}.
\end{array} \right.
\end{equation}
The characteristic ejecta velocity is discribled as
\begin{equation}
V_{\rm ej}=\sqrt \frac{10(k-5)E_{\rm sn}}{3(k-3)M_{\rm ej}}
\end{equation}
here we set the energy of supernova explosion $E_{\rm sn}=1.0\times10^{51}$ erg and the ejecta mass $M_{\rm{ej}}=2M_{\odot}$, $k=9$ is the power-law index of this distribution in consideration of possible supernova type Ib/c and IIb. For the sake of simplicity, the surrounding medium density $n_0$ is assumed to be uniform, its value is selected according to the position of the shock front in consideration of different radial profiles of X-ray and gamma-ray radiations. In our work, two brighter regions (i.e. Region 3 and Region 4 in \citet{H.E.S.S.Collaboration2018}) in both X-ray and gamma-ray bands are singled out as typical cases, we also call them as "Case 1" and "Case 2" for the Region 3 and Region 4 in following sections.

\begin{figure}
	\centering
	\includegraphics[width=15cm, angle=0]{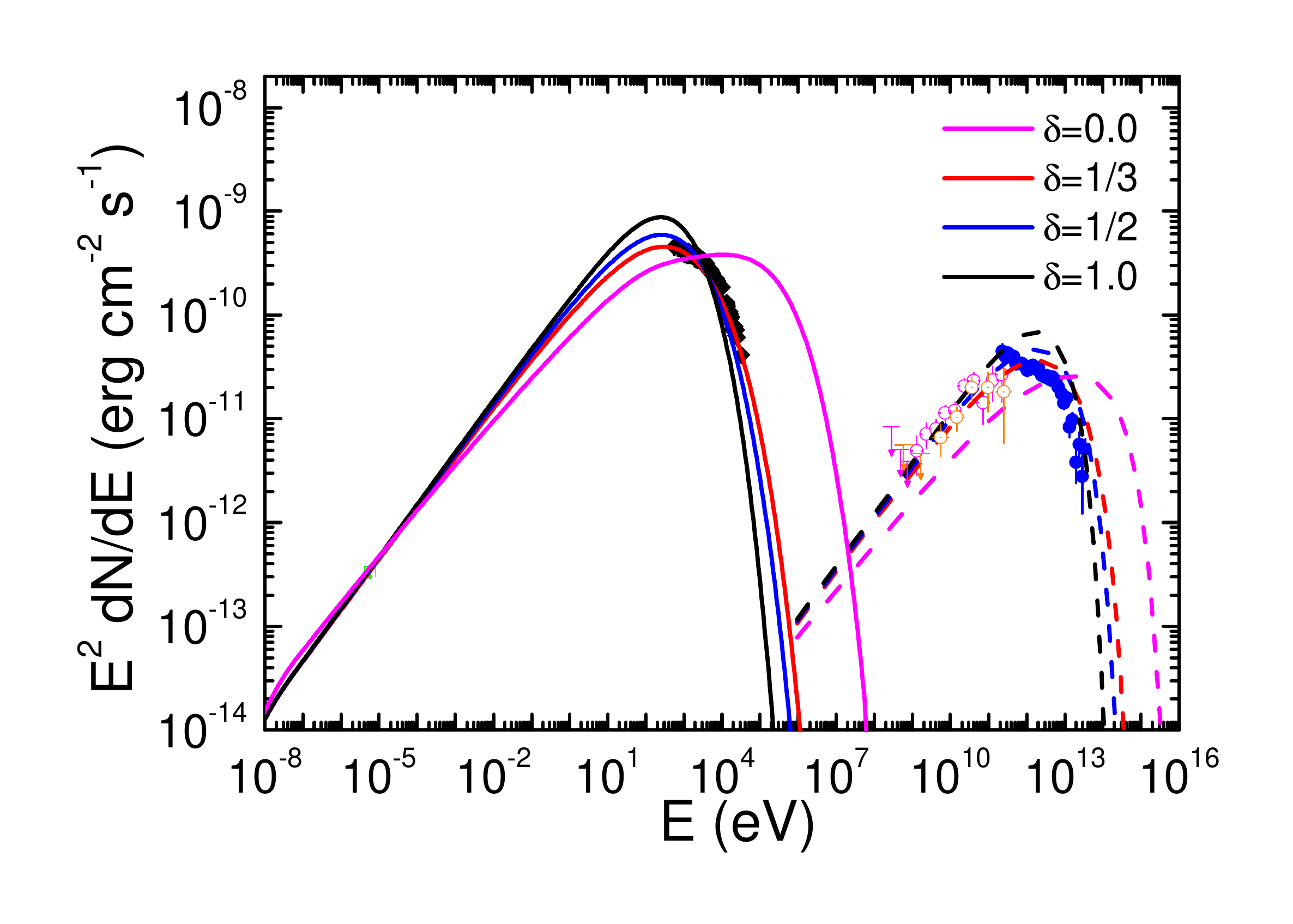}
	\caption{Nonthermal photon spectra of RX J1713.7-3946 with different $\delta$. The value of $\delta$ is equal to 0.0 for the magenta lines with $\chi=4.0\times10^{-3}$, 1/3 for the red lines with $\chi=1.2\times10^{-3}$, 1/2 for the blue lines with $\chi=0.45\times10^{-3}$, and 1.0 for the black lines with $\chi=0.015\times10^{-3}$, the solid and dashed lines represent the synchrotron radiation and the ICS respectively. Observation data from the whole remnant are shown in radio band (green hollow squares from the ATCA, Acero et al. 2009), X-ray bands (black solid squares from the Suzaku, Tanaka et al. 2008), gamma-ray bands (orange hollow circles from the Fermi, Abdo et al. 2011; magenta hollow open circles and blue solid filled circles from the Fermi and H.E.S.S. respectively, H.E.S.S. Collaboration et al. 2018).  }
	\label{fig:1}
\end{figure}

The problem of electron injection is still rarely understood for the diffusion shock acceleration theory in SNRs. A rather high injection energy of electrons 100 MeV is assumed in consideration of the suprathermal electron injection. As discussed in \citet{Zirakashvili2012}, the electron injection efficiency $\eta ^f$ at the forward shock is probably relative to the photo-ionization process of accelerated ions, and $\eta ^f\propto\dot{R}_f^2/c^2$. In what follows, the accelerated electrons are considered as test particles (the electrons are energetically unimportant), and to reproduce the radio fluxes of RX J1713.7-3946, we assume that electrons are injected at the forward shock with a rather high injection energy of $p_f c = 100~{\rm MeV}$ and a low efficiency of $\eta ^f=10^{-7}\dot{R}_f^2/c^2$ that is taken to be independent of the shock velocity, $c$ is the velocity of light. Subsequently, particle diffusion is formulated with above-mentioned four values of $\delta$ by combining with the observation data. Here $B_f=B_0=6~\mu \rm G$, $B_m=45~\mu \rm G$, $\alpha=4$ and $\xi=0.4$ are fixed, we calculated the photon emission spectra of RX J1713.7-3946 shown in Fig.\ref{fig:1}. As you can see from the  Fig.\ref{fig:1}, the emissions of radio to X-ray bands and gamma-ray emissions are respectively from the synchrotron radiation and ICS of relativistic electrons, background photon field of the ICS is composed of cosmic microwave background radiation (CMB) with a temperature of 2.7 K and an energy density of 0.26 eV cm$^{-3}$, and an interstellar infrared radiation field (IIRF) with 30 K and 0.3 eV cm$^{-3}$ \citep[e.g.][]{Porter2006,Finke2012}. As shown in Fig.\ref{fig:1}, if the diffusion is energy-independent, i.e., the index $\delta=0.0$, it seems impossible to well fit the multi-band photon emission spectrum in this situation. In consequence, the energy-dependent diffusion should be a relatively better choice for the sake of fitting the photon emission spectra, especially at $\delta=1/3$, therefore we always adopt $\delta=1/3$ in following calculations.

\begin{figure*}
	\centering
		\begin{minipage}{8cm}
			\includegraphics[width=8.7cm]{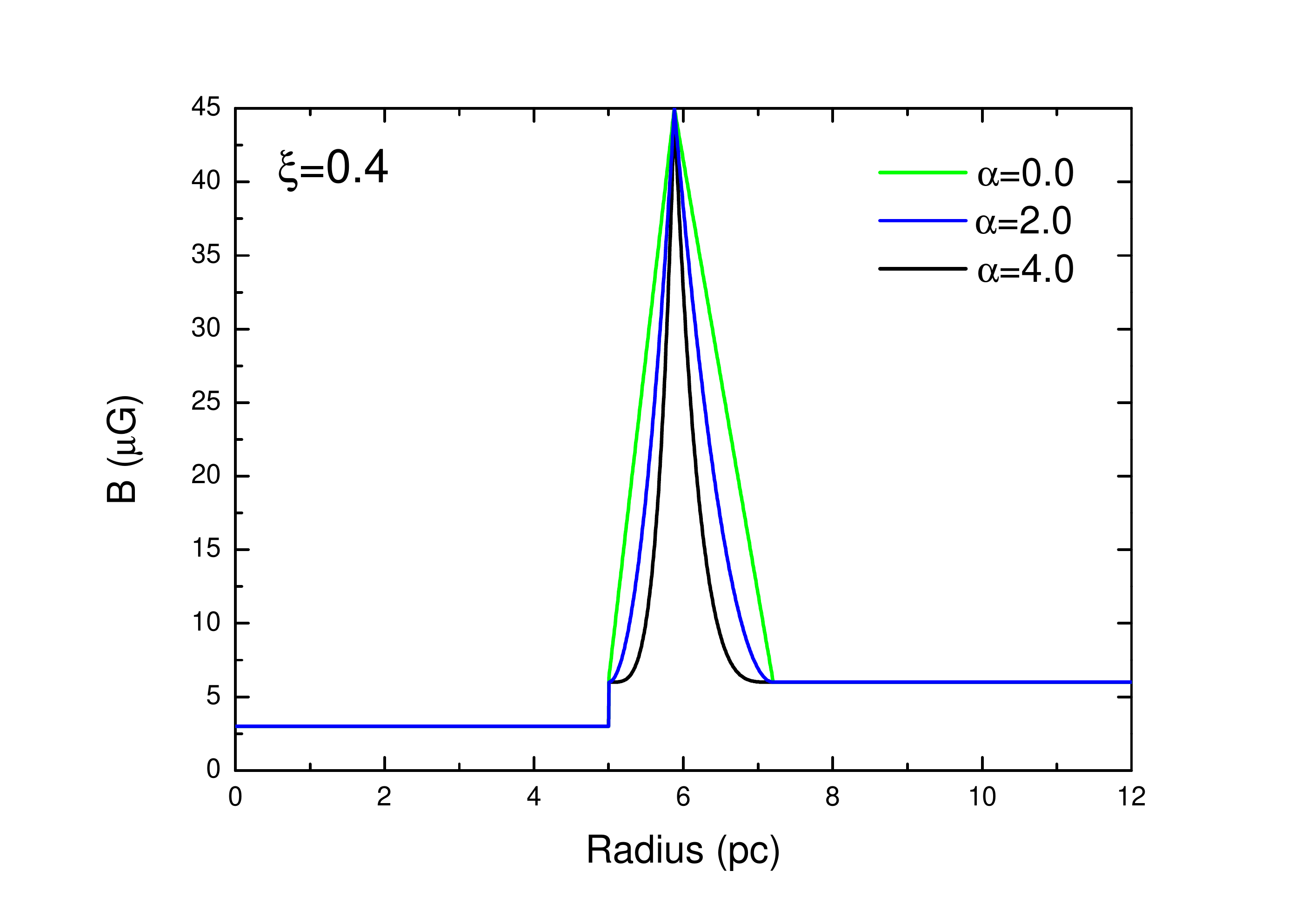}
		\end{minipage}
		\begin{minipage}{8cm}
        	\includegraphics[width=8.7cm]{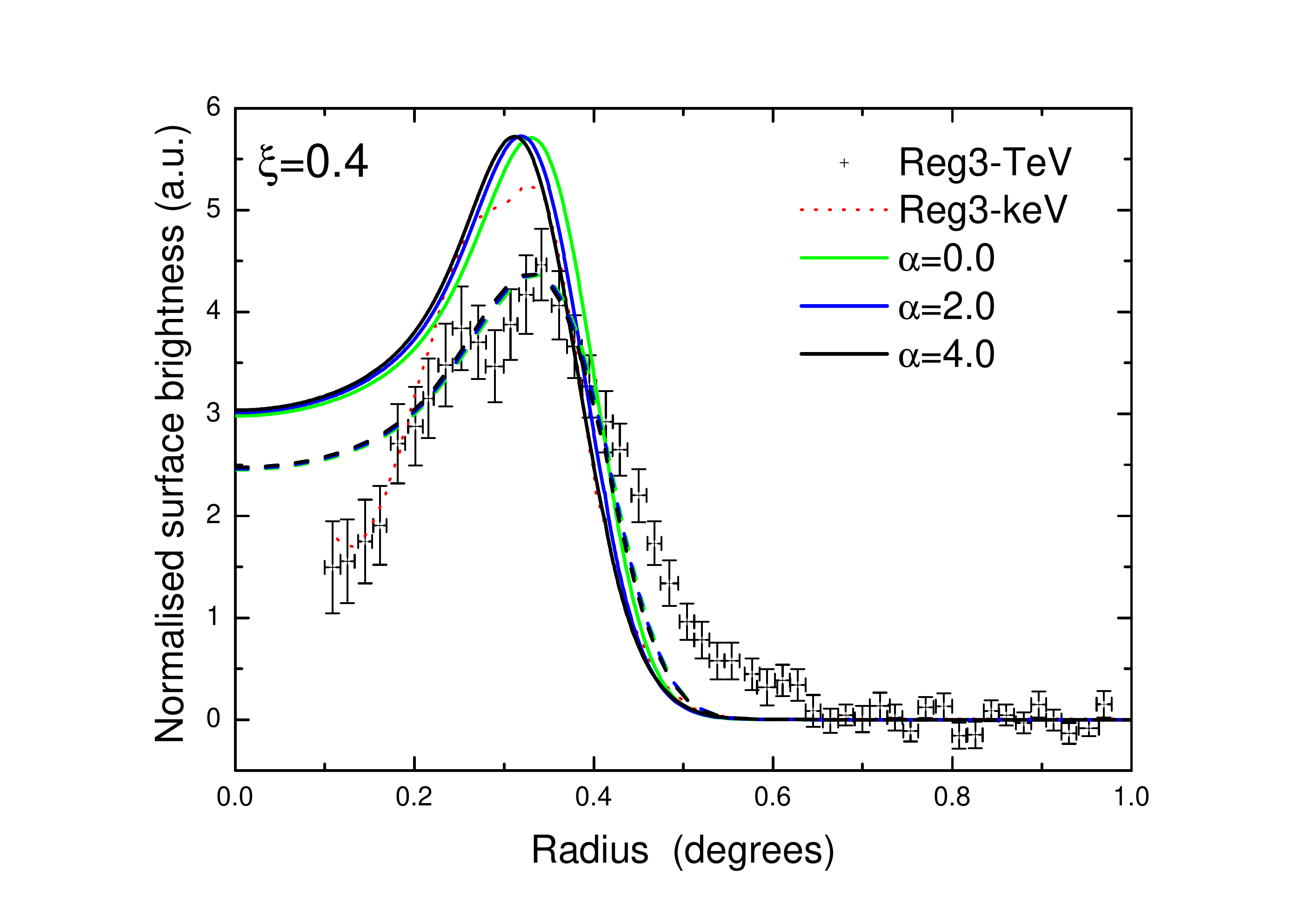}
        \end{minipage}

 		\begin{minipage}{8cm}
        	\includegraphics[width=8.7cm]{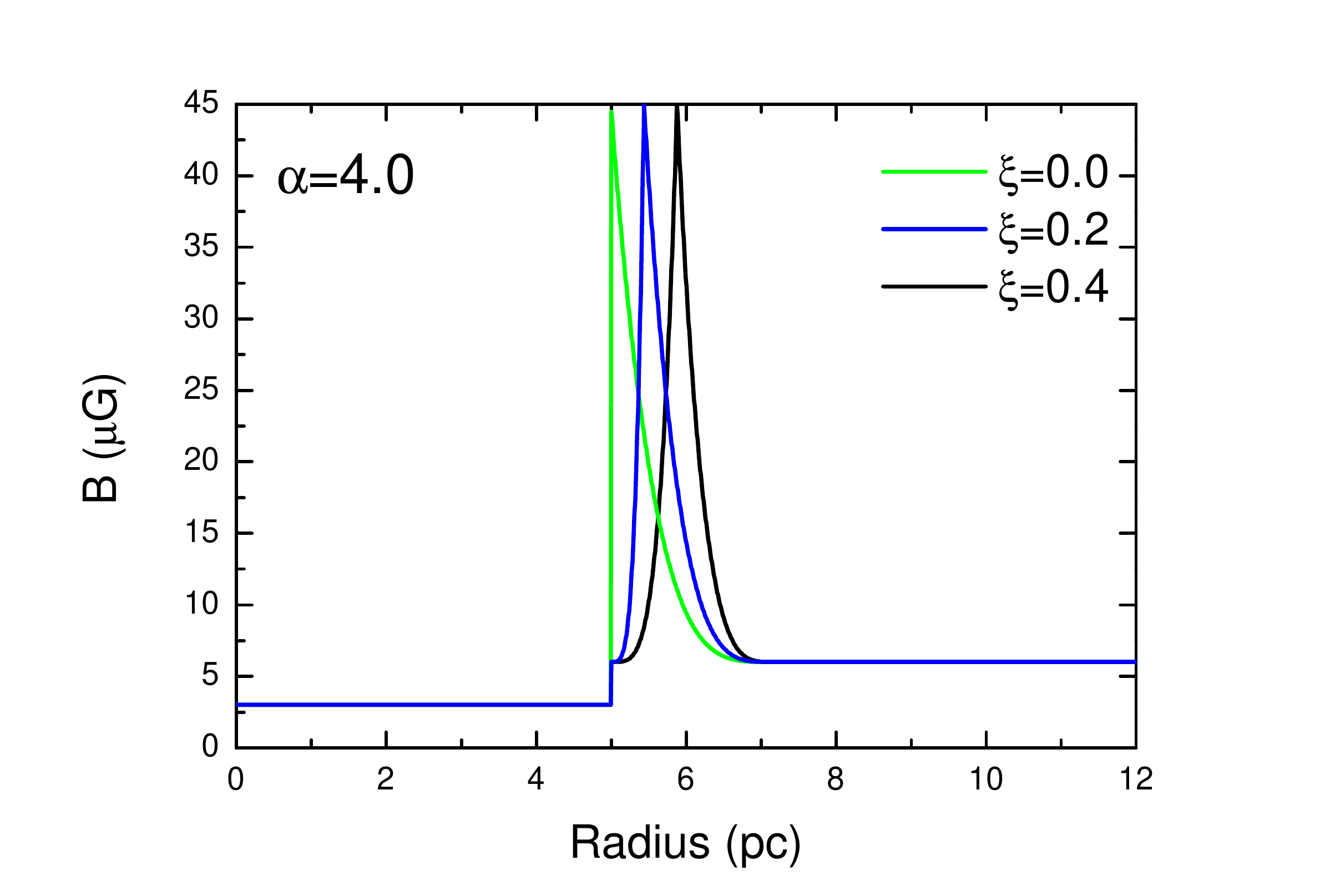}
        \end{minipage}
        \begin{minipage}{8cm}
        	\includegraphics[width=8.7cm]{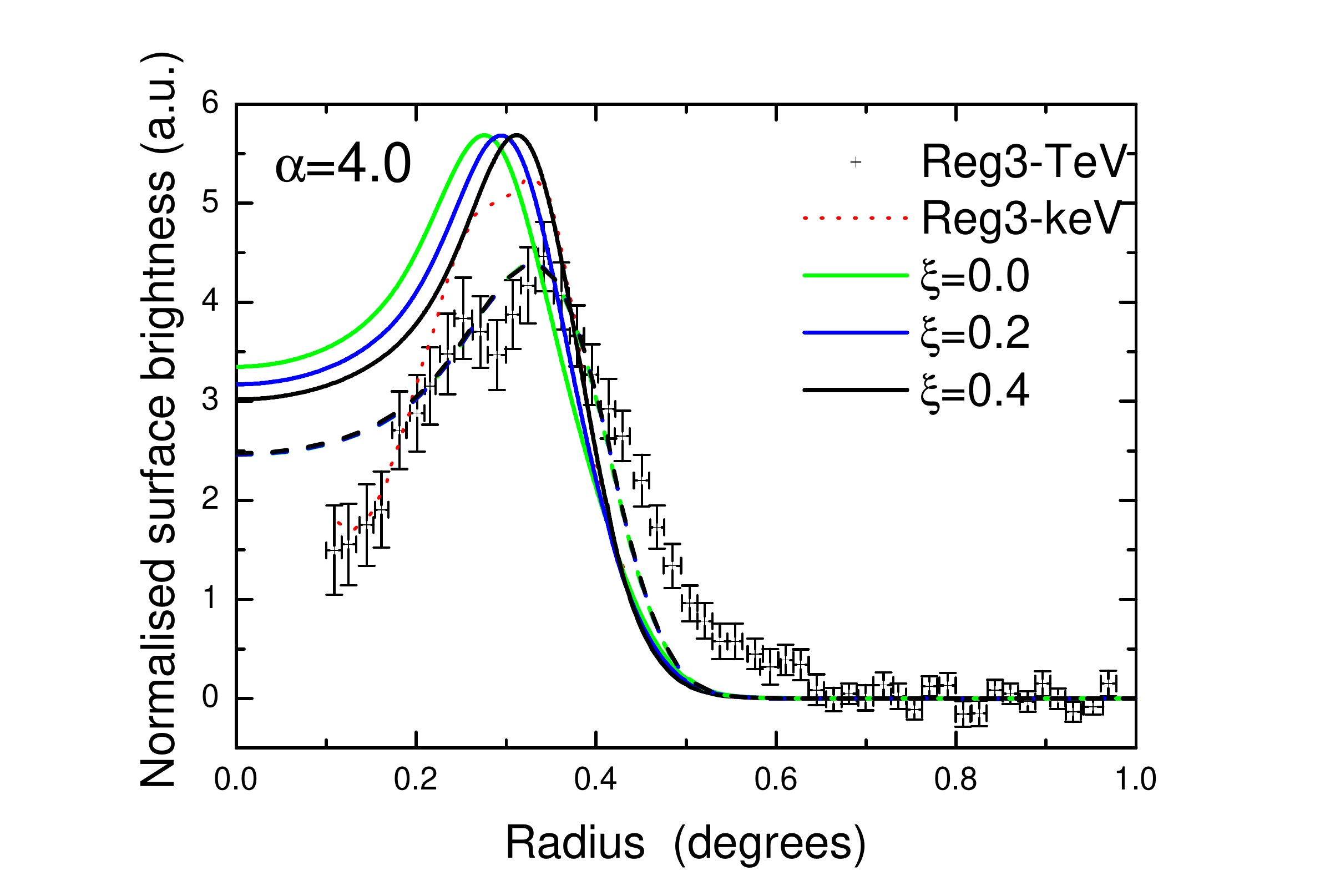}
        \end{minipage}
	\caption{Distributions of magnetic field strength and radial profile in Case 1. Left panel: The distribution of the magnetic field strength varying with radius, $\xi=0.4$ is fixed with $\alpha=1.0, 2.0, 4.0$ in upper Left,  and $\alpha=4.0$ is fixed with $\xi=0.0, 0.2, 0.4$ in lower Left. Right panel: Radial profiles of RX J1713.7-3946 varying with radius corresponding to the left figure in Case 1, the X-ray profile with energy 1-10 keV (red dotted line) is extracted from the XMM-Newton map, gamma-ray profile above 250 GeV (black crosses) is from the H.E.S.S. maps, the same convolution is processed with $0.048^\circ$ \citep{H.E.S.S.Collaboration2018}. Theoretical calculation results are shown as the dashed lines (gamma-ray profiles) and solid lines (X-ray profiles) in right panel.}
	\label{fig:2}
\end{figure*}

In general, the non-thermal X-ray emissions are mainly from the synchrotron radiation of high energy electrons, and the X-ray radiation intensity is proportional to magnetic field strength. In consideration of the effect of the magnetic field on the radial radiation profile in X-ray band, we calculate the radial profiles of RX J1713.7-3946 in X-ray and gamma-ray bands by selecting different parameters $\alpha$ and $\xi$. As shown in Fig.\ref{fig:2}, we provide different distributions of magnetic field strength, and calculate the corresponding radial profiles for Case 1. The gamma-ray radial profile is almost impervious to the distribution change that does not depend on the magnetic field strength because of different radiation mechanism. It's obvious that the result for $\alpha=4.0$ and $\xi=0.4$ is consistent with the observed X-ray radial profile for this case, although there are still some differences in some regions, such as the innermost region and the peak position, more detailed analysis will be carried out in the following sections. It can be seen roughly from the Fig.\ref{fig:2} that $\alpha$ mainly determines the profile shape in the radius range from $0.3^\circ$ to $0.5^\circ$, while the peak position of the profile is affected by $\xi$, which is a crucial factor to fit the observation in the region with radius $< 0.4^\circ$.

\begin{figure*}
	\centering
	\begin{minipage}{8cm}
		\includegraphics[width=8.7cm]{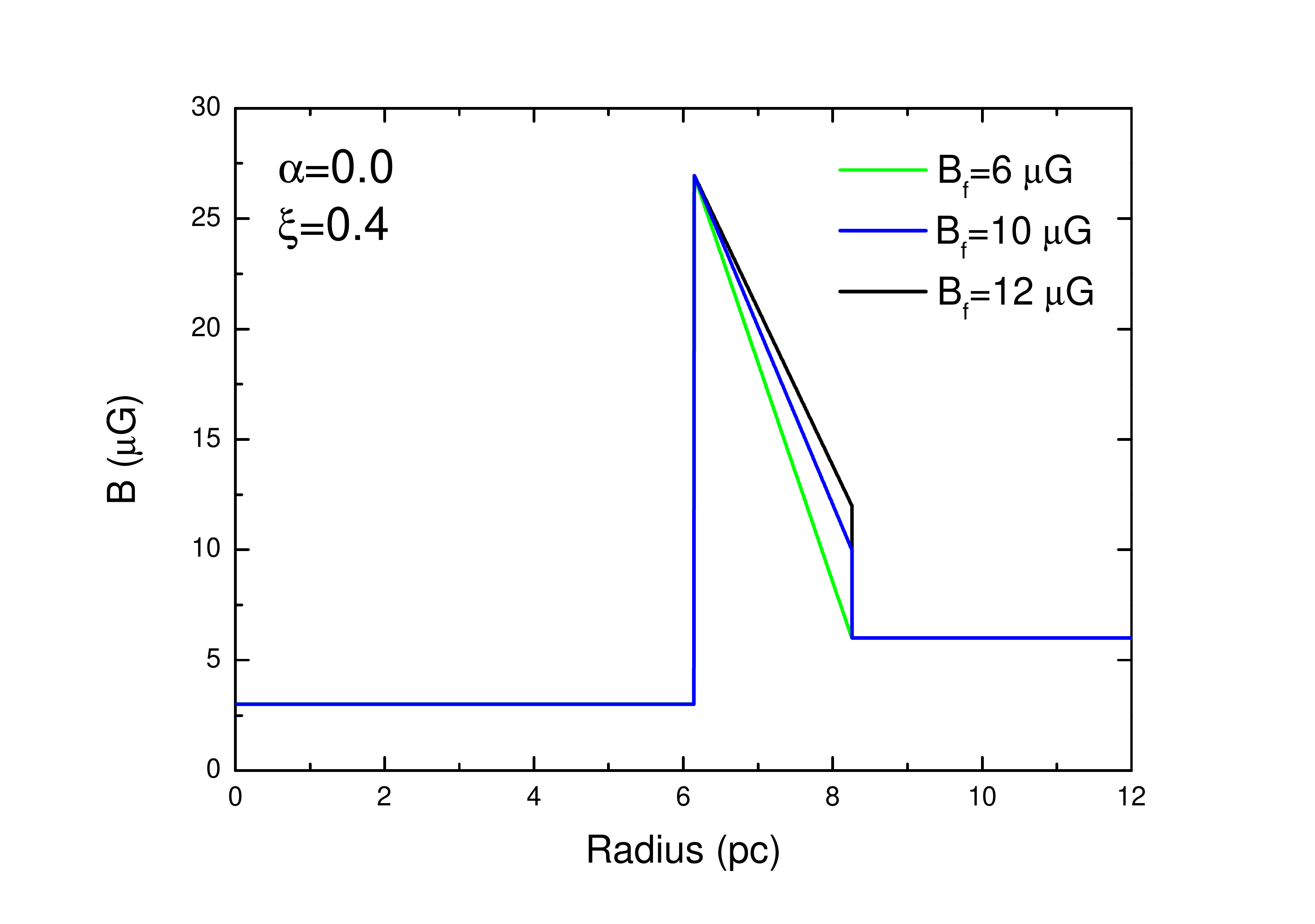}
	\end{minipage}
	\begin{minipage}{8cm}
		\includegraphics[width=8.7cm]{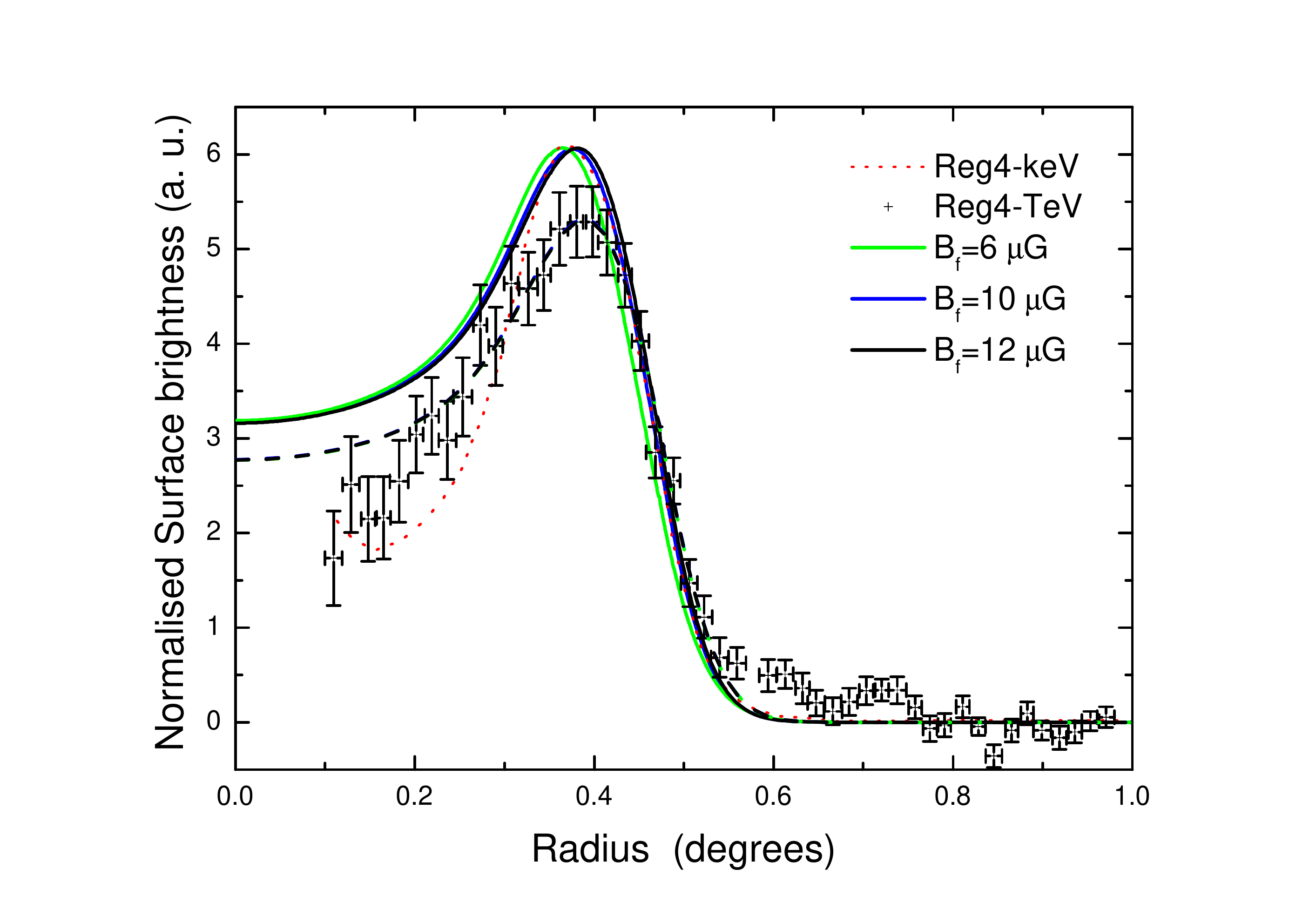}
	\end{minipage}
	\caption{Distributions of magnetic field strength and radial profiles in Case 2. Left panel: The distribution of the magnetic field strength varying with radius,  $\alpha=1.0$ and $\xi=0.0$ are fixed with $B_f=6.0, 10, 12~\mu\rm G$. Right panel: Radial profiles of RX J1713.7-3946 varying with radius corresponding to the left figure, the descriptions of experimental data and theoretical calculation results are described as same as those of Fig\ref{fig:2}, just for different regions. }
	\label{fig:3}
\end{figure*}

Compared to the Case 1, the gamma-ray emission extending beyond the X-ray emitting shell is slightly less obvious in Case 2 where we also adjust the ambient density from 0.45 cm$^{-3}$ for case 1 to 0.2 cm$^{-3}$  to fit the location of the brightness peak, the X-ray Surface brightness drops more slowly with varying position from the peak to the forward shock front in this case, thus the distribution of magnetic field strength should be different from the result of the Case 1, and the value of $\alpha$ should be smaller. Now we set $\xi=0.0$ and $\alpha=1.0$ for a magnetic field strength that is changing linearly with position, $B_0=6~\mu \rm G$, $B_f=6,10,12~\mu \rm G$ are adopted by considering different compression amplification at the forward shock front, $B_m=27~\mu \rm G$ is chosen to fit the muti-band emission spectra of RX J1713.7-3946. Some different results are shown in Fig\ref{fig:3}, it is observed that $B_f=10~\mu\rm G$ is appropriate to fit the radial profile in Case 2, in other word, the compressional amplification of magnetic field strength is likely to exist at the shock front.

\section{The results of modeling RX J1713.7-3946}
On the basis of the parameter analysis in Section \ref{sect:Para}, we select some appropriate model parameters, and further analyze the shock evolutions, magnetic field distributions, muti-band energy spectra and radial profiles of RX J1713.7-3946 towards fore-mentioned two cases. As previously discussed, here we will take an energy-dependent diffusion with the index $\delta=1/3$ and a position-dependent magnetic field, the parameters $E_{\rm sn}$, $M_{\rm{ej}}$, $k$, $p_{f}$, $\eta ^f$, $\delta$ and $B_0$ are fixed as same as those values determined in Section \ref{sect:Para} for all two cases, while the parameters $n_0$, $\chi$ and other parameters ($B_f$, $B_m$, $\alpha$ and $\xi$) that are used to describe the magnetic field strength are discrepant in two different cases, as listed in Table \ref{tab1}.
\begin{table}
    \centering
    \caption{Model Parameters for different cases.}
	\begin{tabular}{ccccccc}
		\hline
		Case & $n_0$(cm$^{-3}$)  &$\chi (10^{-3})$   &$\alpha$     &$\xi$      &$B_f (\mu \rm G)$     &$B_m(\mu \rm G)$        \\
		\hline
		\hline
		Case 1      &$0.45$      &$1.15$       &$4.0$          &$0.4$      &$B_0$   &$48$               \\
		Case 2      &$0.20$      &$1.5$        &$1.0$          &$0.0$      &$10.0$  &$25$               \\
		\hline
	\end{tabular}
	\label{tab1}
\end{table}

\begin{figure*}
	\centering
	\begin{minipage}{8cm}
		\includegraphics[width=8.7cm]{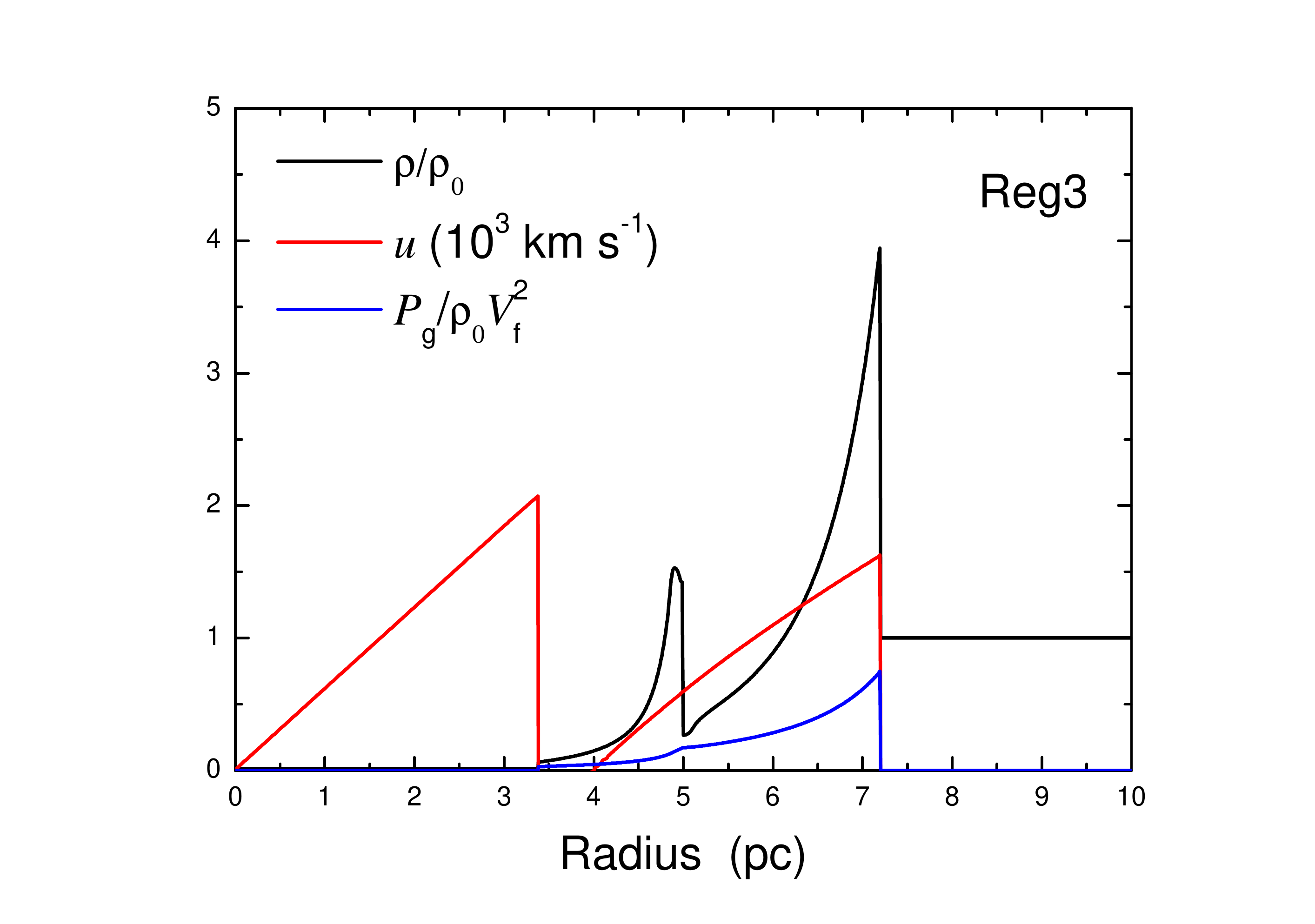}
	\end{minipage}
	\begin{minipage}{8cm}
		\includegraphics[width=8.7cm]{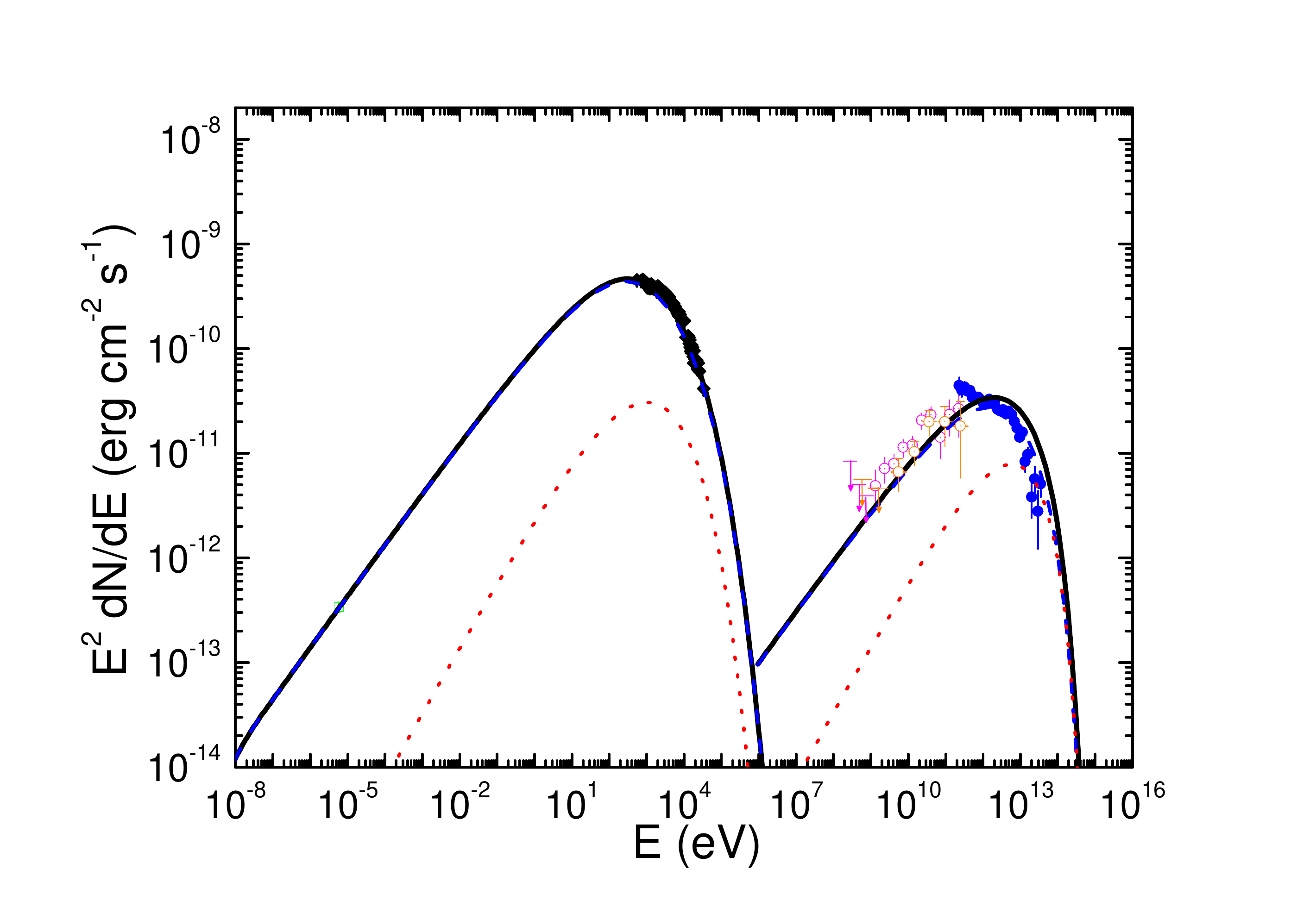}
	\end{minipage}
	
	\begin{minipage}{8cm}
		\includegraphics[width=8.7cm]{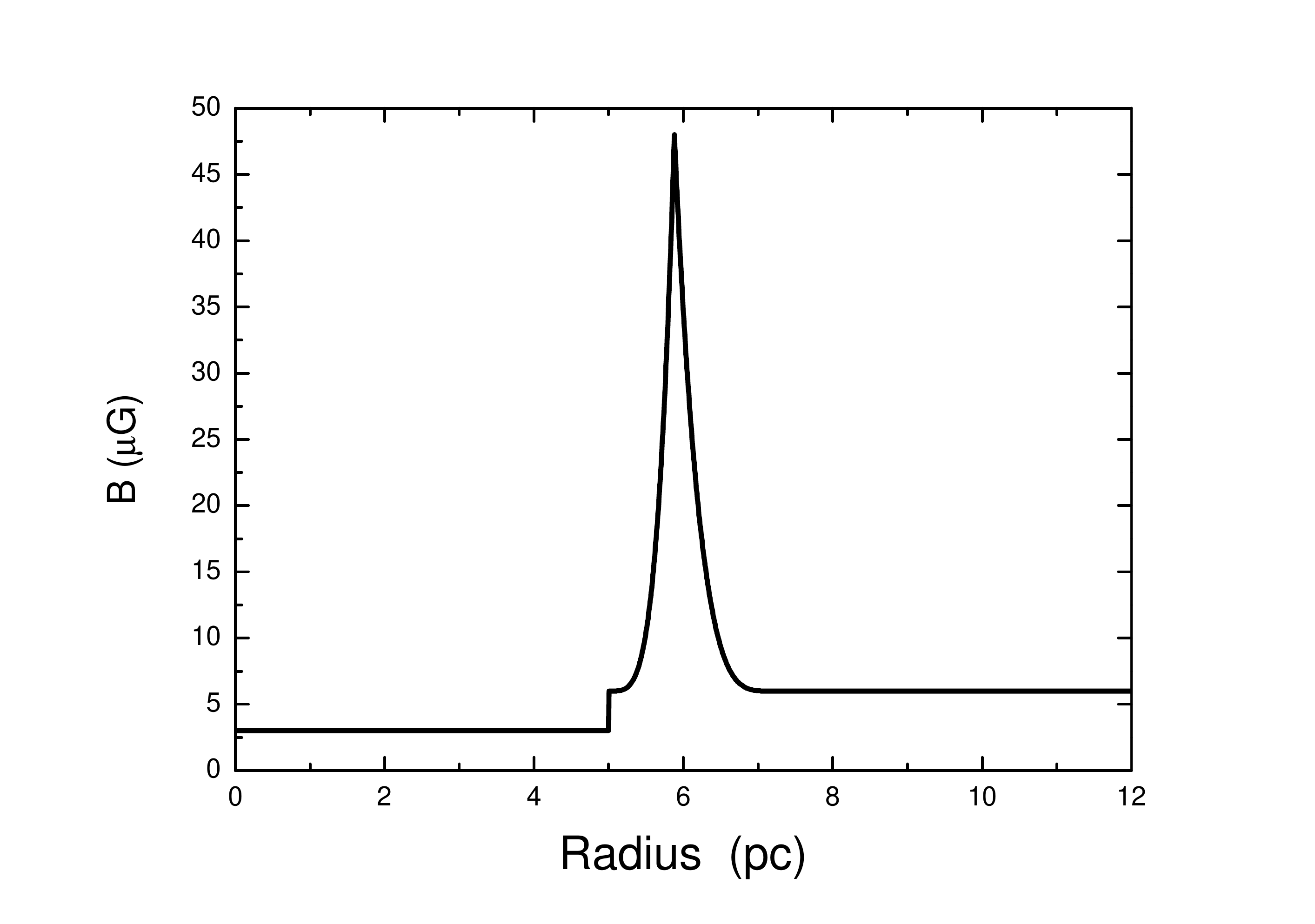}
	\end{minipage}
	\begin{minipage}{8cm}
		\includegraphics[width=8.7cm]{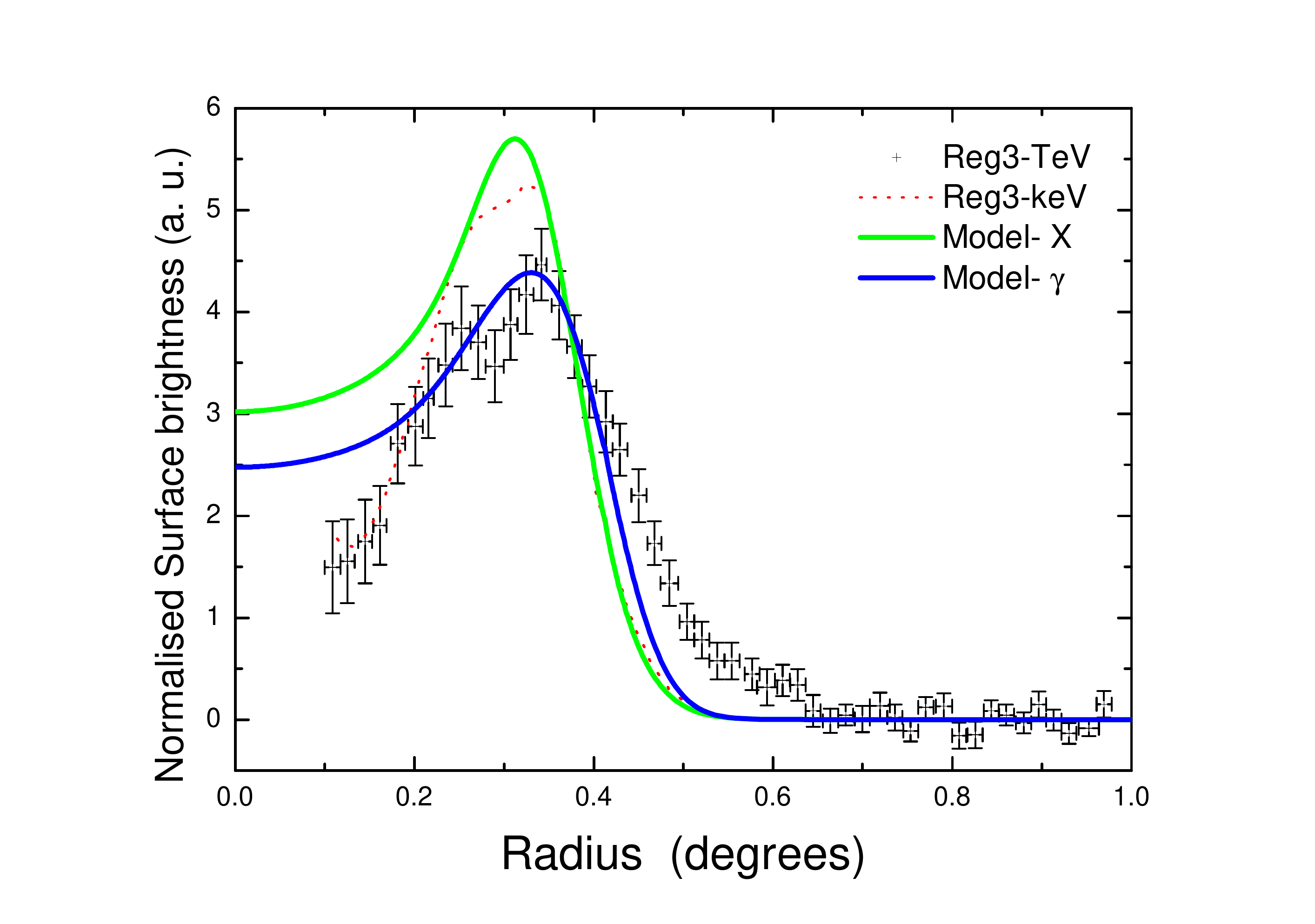}
	\end{minipage}
	\caption{Nonthermal photon spectrum and radial dependencies of gas characteristics, magnetic field strength and surface brightness in Case 1. Upper Left panel: Radial dependencies of the gas density with unit $\rho_0$ (black solid line), the gas velocity with unit $10^3$ km s$^{-1}$ (red solid line), and the gas pressure $\rho_0V_f^2$ (blue solid line). Upper right panel: Nonthermal spectrum of RX J1713.7-3946, the emission from downstream of forward shock is represented as the blue dashed lines, the emission from upstream for the red dotted lines, and the black lines represent the total radiation.
    Lower Left panel: Radial distribution of magnetic field strength. Lower right panel: Radial profiles of photon emssions in X-ray bands (green solid line) and gamma-ray band (blue solid line).	
    }
	\label{fig:4}
\end{figure*}

\begin{figure*}
	\centering
	\begin{minipage}{8cm}
		\includegraphics[width=8.7cm]{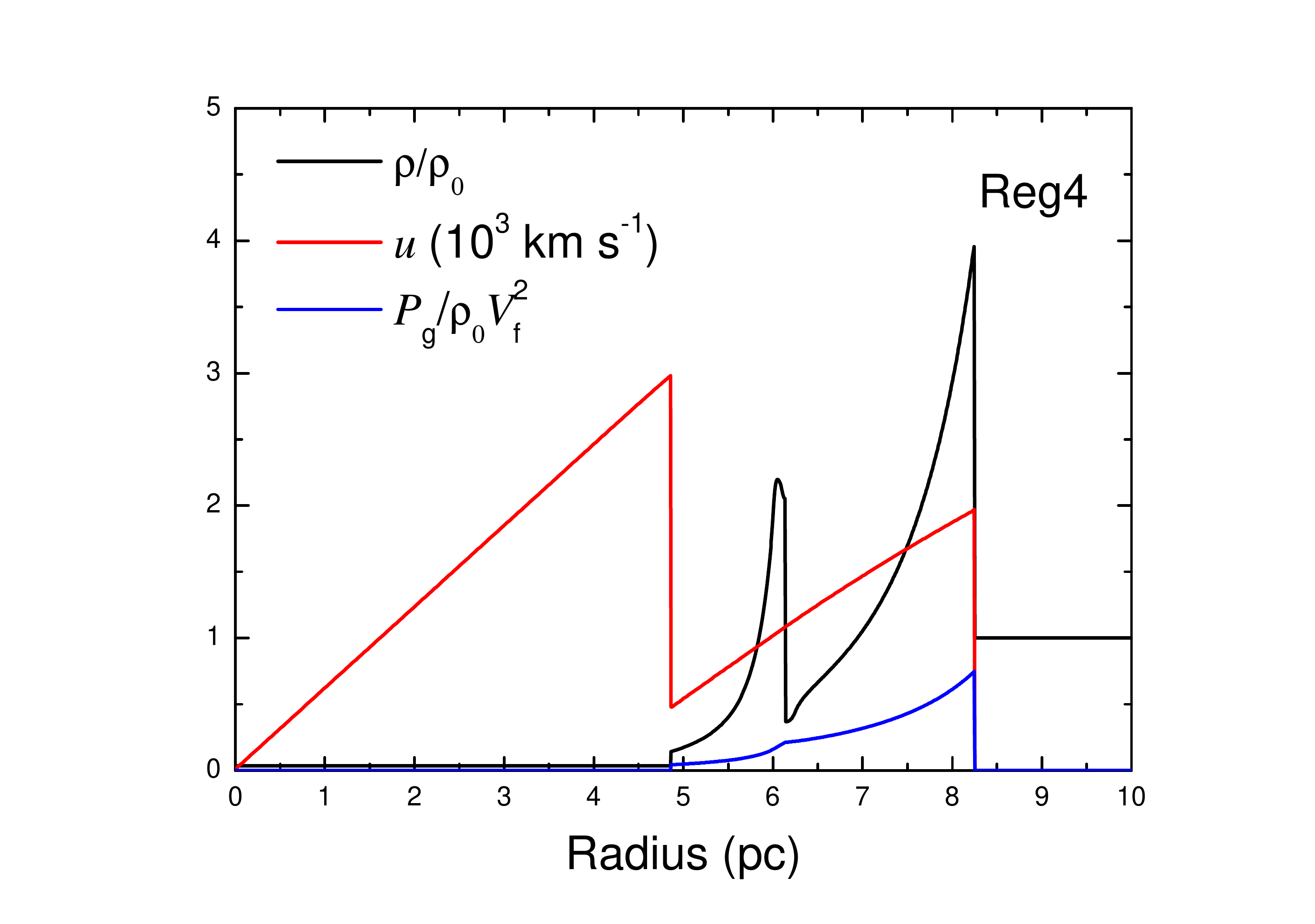}
	\end{minipage}
	\begin{minipage}{8cm}
		\includegraphics[width=8.7cm]{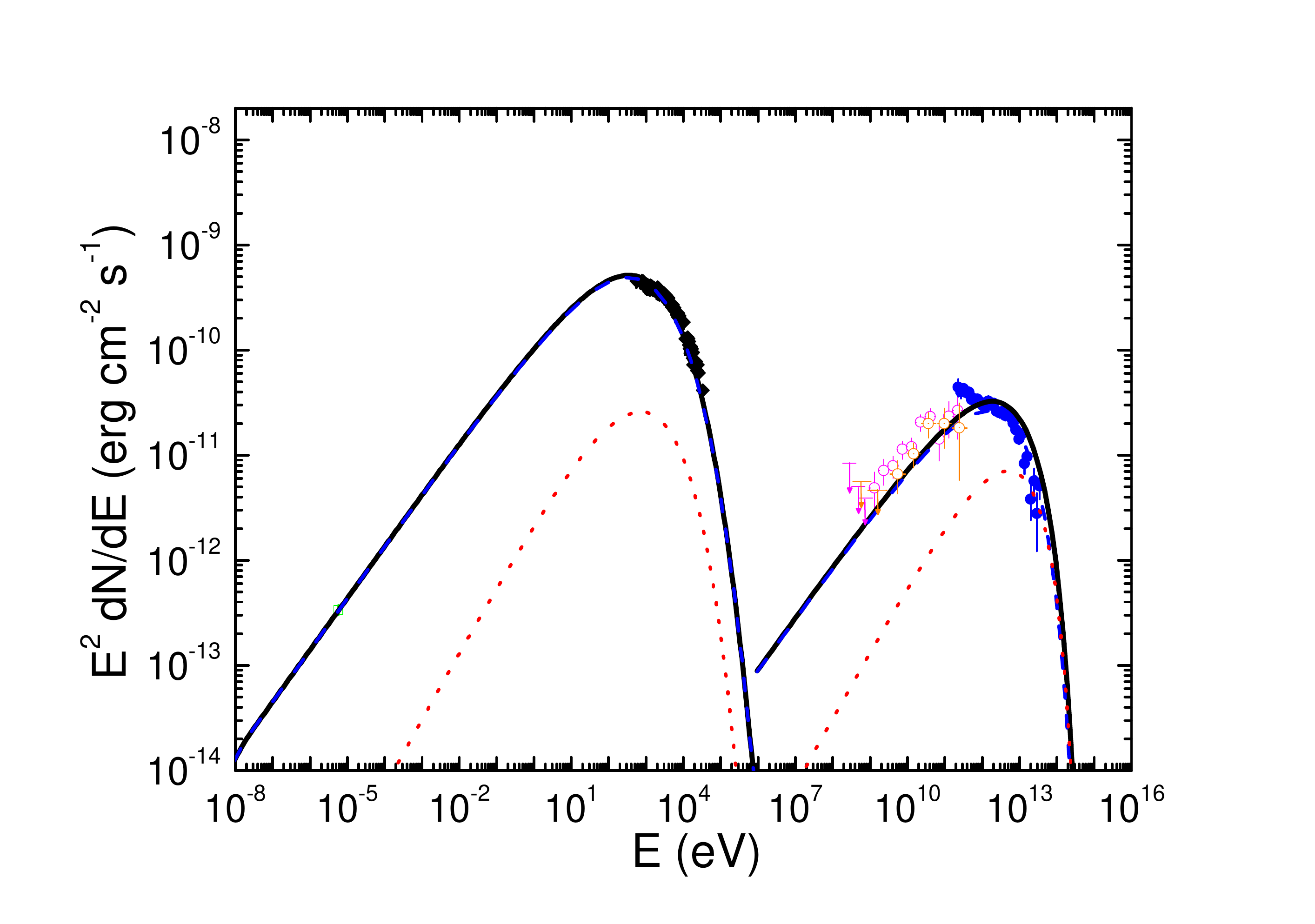}
	\end{minipage}
	
	\begin{minipage}{8cm}
		\includegraphics[width=8.7cm]{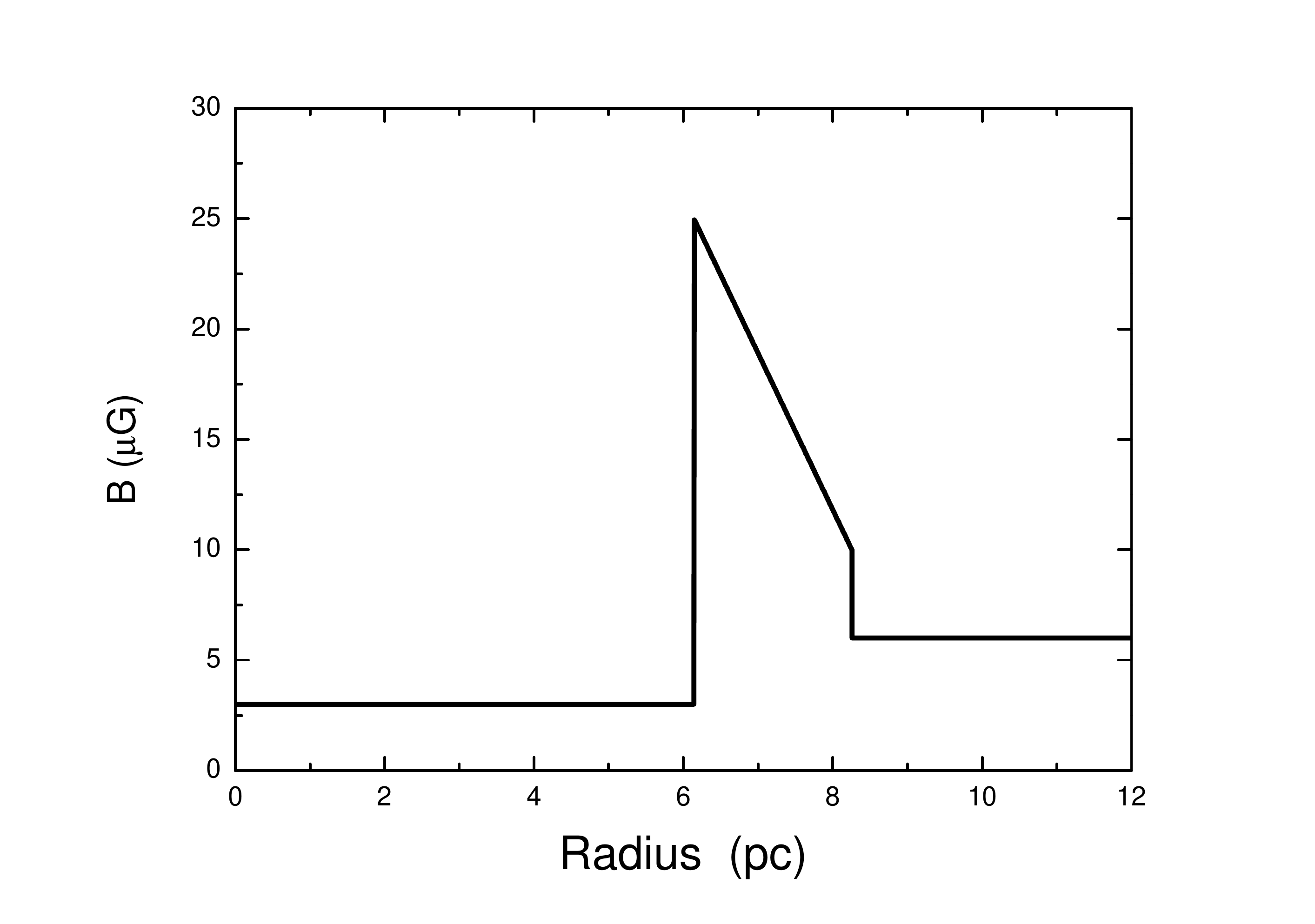}
	\end{minipage}
	\begin{minipage}{8cm}
		\includegraphics[width=8.7cm]{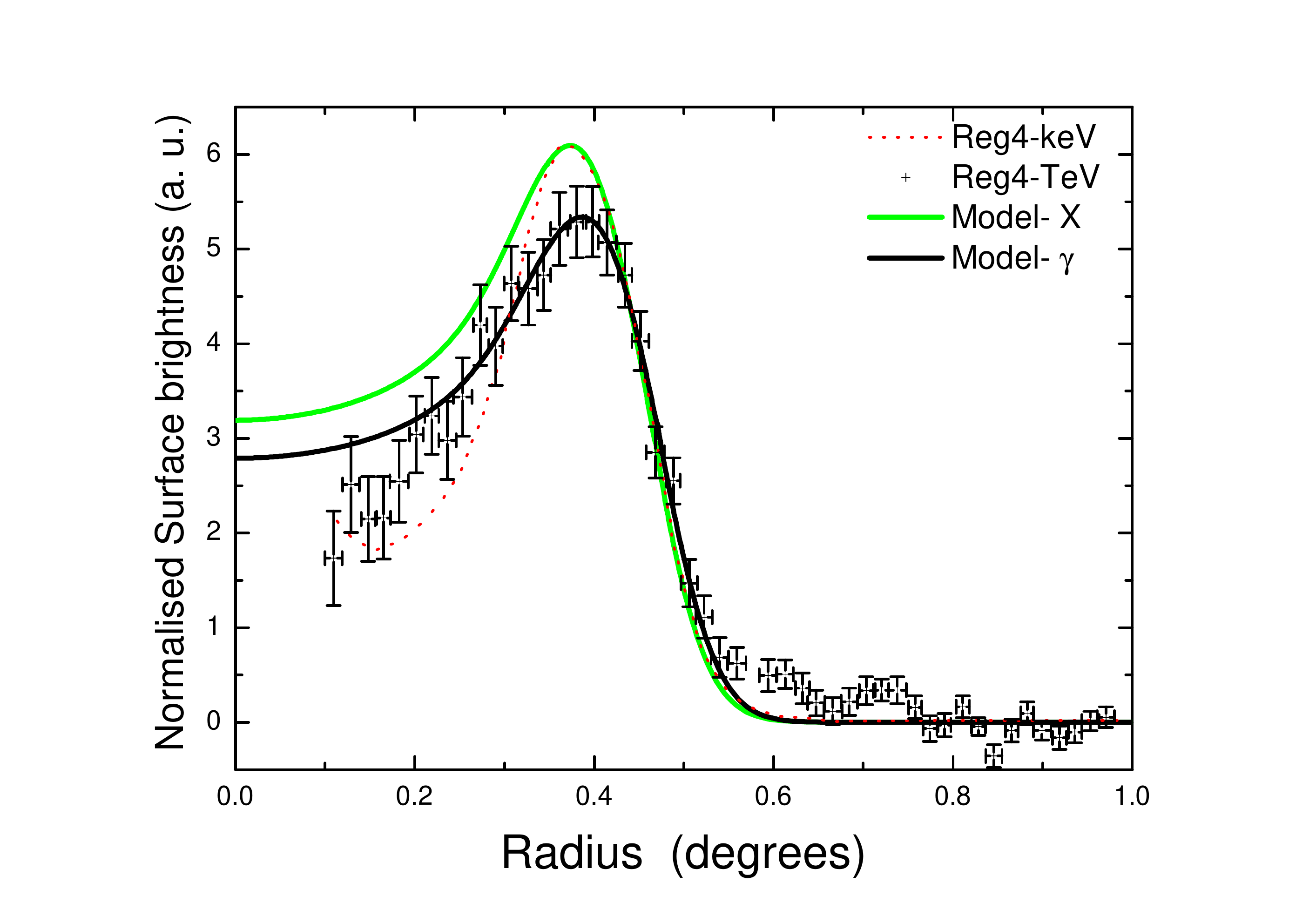}
	\end{minipage}
	\caption{Nonthermal photon spectrum and radial dependencies of gas characteristics, magnetic field strength and surface brightness in Case 2. The description of this figure is the same as Fig.\ref{fig:4}. All observational data have also been depicted in Fig.\ref{fig:1} and Fig.\ref{fig:3}}
	\label{fig:5}
\end{figure*}

In this section, we apply the model to explore the radial dependencies of RX J1713.7-3946 with gas characteristics, magnetic field strength, and further study the multi-band nonthermal emission and surface brightness distribution for this remnant. For two different cases, with the age ($t_{\rm age}=1625~\rm {yr}$) of the remnant, we can inversely derive the surrounding medium density $n_0$ via the observed brightness distributions after setting initial and boundary conditions. As you can see in Table \ref{tab1}, the parameters associated with the magnetic field strength are determined by the shape of the radial profiles of X-ray emissions. Although the maximum value of the magnetic field strength is different for two cases, the average values of them seem to be close, which should be determined by the observed X-ray to gamma-ray energy flux ratio when gamma emission of RX J1713.7-3946 is from the ICS of relativistic electrons. In Fig.\ref{fig:4} and Fig.\ref{fig:5}, the positions of the contact discontinuity (between the ejecta and the interstellar gas), the forward and reverse shocks can be seen in the upper left panel, their radius are respectively located at $R_c=5.0~\rm{pc}$, $R_f=7.2~\rm{pc}$ and $R_b=3.38~\rm{pc}$ for Case 1, while $R_c=6.14~\rm{pc}$, $R_f=8.25~\rm{pc}$ and $R_b=4.86~\rm{pc}$ for Case 2. $R_f=7.2~\rm{pc}$ for Case 1 is close to the result of $\sim7~\rm{pc}$ that was given by the model fit in \citet{Zhang2016}. The speed of the forward shock $V_f$ is equal to $2170~\rm {km~s^{-1}}$ and $2630~\rm {km~s^{-1}}$ for Case 1 and Case2 respectively, which are much less than the upper limit of the shock speed $4500~\rm {km~s^{-1}}$ \citep{Uchiyama2007}. Recently, the shock wave speed at the northwestern shell is measured to be $3900\pm300~\rm {km~s^{-1}}$ with an estimated distance of d$=1~kpc$\citep{Tsuji2016}, \citet{Tanaka2020} also gave the measured velocities are $3800\pm100~\rm {km~s^{-1}}$ and $2300\pm200~\rm {km~s^{-1}}$ for two different X-ray bright blobs at the western edge of the shell, respectively.

For mutiband emission spectra of RX J1713.7-3946 in upper right panel of Fig.\ref{fig:4} and Fig.\ref{fig:5}, the emissions from radio to X-ray bands are contributed by the synchrotron emission of relativistic electrons, the GeV-TeV gamma-ray emissions are from the ICS of these electrons on the background photon fields, although the fitted results are not perfect. It is obvious that the contribution from the downstream of forward shock is dominant over the total emissions, while the hard X-ray and TeV gamma-ray emissions could be partly from the shock upstream. The structure of magnetic field in the shock downstream is closely related to the X-ray radiation profile. \citet{Xu2016} investigated the magnetic field amplification in the context of fully ionized and weakly ionized gas, and found that the nonlinear dynamo in fully ionized gas could led to a linear-in-time growth of the magnetic energy by means of the turbulent magnetic diffusion (i.e., the growth of magnetic field is non-linear with time), while the dynamo was characterized by a linear-in-time growth of magnetic field in the weakly ionized medium. In our calculations, for the sake of fiting the observed data, the growth of magnetic field strength is simply assumed to be nonlinear and linear with position for Case 1 and Case 2 respectively in shock downstream, a possible compression amplification at the shock front is also taken into account to better fit the observations in Case 2.

From the lower right panel we can see that radial profiles with energy above 250 GeV are well consistent with the corresponding experimental data around the peak position of surface brightness for two cases, but there are still differences in innermost regions with radius ($r<0.2^\circ$) and outermost regions ($r>0.4^\circ$ for Case 1, $r<0.5^\circ$ for Case 2), the theoretical values are on the high side in the innermost regions, on the contrary, theoretical values are on the low side in the outermost regions. In both cases, radial profiles with energy above 1-10 keV are in good agreement with the experimental data outside the peak position, but theoretical values are significantly higher than the observed values inside the peak position. Certainly, these biases are understandable because of the spherically symmetric hypothesis in our theoretical model, as a matter of fact, the local physical environment could be very complex, such as anisotropic surrounding medium density, partially or fully ionized local gas and so on. In the case of non-spherical symmetry, theoretical results in the outer layer of the remnant should be closer to our spherically symmetric results on account of the local projection effect, therefore it is reasonable to explain X-ray radial profiles. The gamma-ray emission extending beyond the X-ray emitting shell is not obvious in our results, especially in the outer regions, where gamma-ray emissions could be provided by the highest-energy particles that escaped from the shock via interacting with the surrounding medium \citep[e.g.][]{Li2010,Li2012,Ohira2011,Ohira2012,Yang2015b,Zhang2016}.

\section{Summary and Discussion}
We use the DSA model with test particle approximation, by considering energy-dependent diffusion and position-dependent magnetic field strength, to investigate the multi-band photon emission spectra and the radial surface brightness distributions of RX J1713.7-3946 in the leptonic scenario for the $\gamma$-ray emission. Here we have ignored the contribution from reverse shock because of its more uncertainty. After selecting the appropriate parameters, the equations for hydrodynamic and particle propagation were solved, and then we explored the dynamic evolution of RX J1713.7-3946 and got the electron momentum distributions, photon emission spectra and radial profiles in two different zones. In the calculation, we have assumed spherical symmetry for simplify, but for real SNR, and there should be different mean density in different regions. The diffusion coefficient and magnetic field strength are critical to the results, the diffusion coefficients were both assumed to be energy-dependent with our spectral fit gives the index $\delta=1/3$. For Case 1, the magnetic field strength increased nonlinearly away from the shock front in the downstream region and has achieved the maximum value before reaching the contact discontinuity by possible turbulent amplification, then decreased nonlinearly. For Case 2, It was postulated that there was a compression amplification of the magnetic field strength at the shock front, afterward it increased linearly up to the contact discontinuity. Some detailed theory about magnetic field amplification have also been discussed by many authors \citep[e.g.][]{Bell2004,Giacalone2007,Iapichino2012,Guo2012,Ji2016,Xu2016,Xu2017}, here we have only given some results from the perspective of fitting observations with simple assumption, further detailed theoretical studies still seem to be necessary for real SNR environment, which is beyond the scope of this paper.

The multi-band photon spectra were reproduced by the sychrotron radiation and the ICS from relativistic electrons, but the origin of gamma-ray emission from RX J1713.7-3946 is still an open problem. In particular our model under-estimates the $\gamma$-ray brightness just ahead of the X-ray shell while over-produces X-ray and $\gamma$-ray emission in the central region. \citet{Zhang2016} put forward that the gamma-ray emission could be from the ICS of relativistic electrons, the interaction between the trapped energetic protons and the shocked clumps in the inside of the SNR, or from the diffusive protons that escaped from the shock, and they believed that the outstretched gamma-ray emission should be from hadronic components in outer emitting region. The very hard energy spectrum at low energies could be also from hadronic processes if the remnant was expanding inside a clumpy medium \citep[e.g.][]{Inoue2012,Gabici2014,Celli2019}. Based on the steady state or time-dependent nonlinear DSA model, the hydrodynamic process and the multi-band photon spectra of RX J1713.7-3946 have been studied \citep[e.g.][]{Zirakashvili2010,Ellison2010,Ellison2012,Yasuda2019}. In particular, with considering both hadronic and leptonic scenarios, \citet{Zirakashvili2010} disscussed the radial profiles of brightness distributions of X-ray, gamma-ray, and radio emissions, and fitted the azimuthally averaged TeV gamma-ray radial profile detected by the H.E.S.S.. \citet{Yang2015a} have also studied the azimuthally averaged TeV gamma-ray radial profile of RX J1713.7-3946 with 2D magnetohydrodynamic simulations with leptonic scenario, in their opinion, an lower surrounding medium density could lead to a sharper drop of the TeV brightness near the outer emitting region, and a relatively good fit was presented with the surrounding medium density $n_0=0.8~\rm{cm}^2$, which is even larger than the values used in our model. By using the newer observation data (as same as those in Case 1), \citet{Ohira2017} explained the broken power-law spectrum of GeV-TeV gamma-rays by the ICS and believed that the extended component of gamma-ray profile should be from cosmic-ray precursor of accelerated electrons, they assumed that the magnetic field strength was constant, the diffusion coefficient around the shock front has a radial dependence and was spatially constant in the shock upstream region, but gamma-ray profile in the inner region and the observed X-ray profile can not be well described in their model. Recently, \citet{Acero2019} indicated that the correlation between the X-ray emissions and gamma rays is non-linear with the flux approximately $F_\gamma \propto \sqrt{F_X}$, and they did not confirm that the gamma-ray emissions extend further than the X-ray emissions.

In short, there are still some difficulties to well explain the origin of multi-band emission and radial profiles of surface brightness for RX J1713.7-946 at the same time, future observations with higher sensitivity and angular resolution will be worth looking forward. The morphological and spectral properties of several other supernova remnants have also been detected, such as RX J0852-622, RCW 86 and HESS J1731-347, the non-thermal radiation characteristics from these remnants are very similar to RX J1713.7-3946. Future observations with higher sensitivity and angular resolution, just as the Cherenkov Telescope Array (CTA, \citet{CTA2019}) and the Large High Altitude Air Shower Observatory (LHAASO, \citet{LHAASO2019}) experiments, will provide us more detailed information about gamma-ray emissions from SNRs, and help us to understand the particle acceleration mechanism and distinguish whether the extended component of gamma-ray profile is from the shock precursor or escaping cosmic rays.

\acknowledgments
\section*{acknowledgments}
 We thank Felix Aharonian and Siyao Xu for helpful discussions and insightful comments on a draft. This work is partially supported by National Key R\&D Program of China (2018YFA0404203), the National Science Foundation of China (U1931204, 11303012), the Natural Science Foundation of Yunnan province of China (2018FB011), the Yunnan Talent Program for Young Top-Notch Talents.

\clearpage

\end{document}